\documentclass[prl,twocolumn]{revtex4}
\usepackage{graphicx}
\usepackage{amssymb,amsmath}
\usepackage{enumerate}

\newcommand{\be}{\begin{eqnarray}}
\newcommand{\ee}{\end{eqnarray}}

\begin{document}

\title{Hamiltonian distributed chaos in Arctic and Antarctic Oscillations}

\author{A. Bershadskii}

\affiliation{
ICAR, P.O. Box 31155, Jerusalem 91000, Israel
}

\begin{abstract}
The Arctic and Antarctic Oscillations (AO and AAO indices) are studied using the Hamiltonian distributed chaos approach. Using the daily data (AO since 1950y and AAO since 1979y) it is shown that the power spectra of the both AO and AAO indices exhibit the stretched exponential behaviour  $E(f) \propto \exp-(f/f_0)^{3/4}$ corresponding to the Hamiltonian distributed chaos. The characteristic time scale for the both indices $T_0=1/f_0\simeq 41$ day corresponds to the well 
known from the numerous extratropic observations near 40 day period.

\end{abstract}

\maketitle

  The Arctic Oscillation (AO) and the Antarctic Oscillation (AAO) are primary annular modes of extratropical atmospheric circulation in the Northern and Southern Hemispheres respectively \cite{tw} (see for a comprehensive review Ref. \cite{w} and references therein). In a certain sense these patterns represent a measure of the pressure gradient between the subpolar and polar regions. Figure 1, for instance, shows the loading pattern of the Antarctic Oscillation (AAO), which is the Leading Empirical Orthogonal Function of monthly mean 700 hPa height for 1979-2000yy period \cite{ao}. The daily AAO has been constructed by projecting the daily (700mb height) anomalies poleward of $20^{o}$S onto the shown in the Fig. 1 loading pattern. The AO index has been constructed analogously with replacement of the $20^{o}$S by the $20^{o}$N and the 700mb height by the 1000mb height \cite{ao}. \\
  
   The AO and AAO are characterized by winds circulating around the Arctic and Antarctic. When the AO and AAO indices are in their positive phase, these strong winds confine colder air in the corresponding polar regions. When the indices are in their negative phase the ring of winds becomes weaker, that allows a stronger penetration of colder airmasses into subpolar regions and even to the mid-latitudes. The annular modes exist in the troposphere at all seasons. The AO index is most clearly defined during the winter and early spring (see Fig. 2, for instance). The annular pattern is a nearly zonally symmetric in the middle stratosphere corresponding to the stratospheric polar vortex \cite{bd}. It is believed that AO and AAO play a very important role in the climate variability on scales of seasons and longer. Together with the intertropical regions they can also play a significant role in the very long-term climatic variations \cite{b}. \\
\begin{figure}
\begin{center}
\includegraphics[width=8cm \vspace{-0.3cm}]{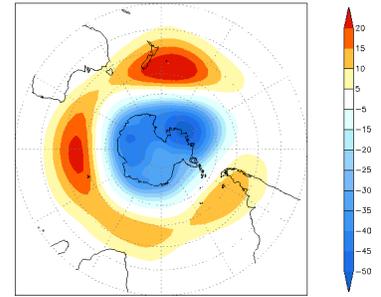}\vspace{-5cm}
\caption{\label{fig1} The Antarctic Leading Empirical Orthogonal Function as a regression map of 700mb height (m) \cite{ao} } 
\end{center}
\end{figure}

\begin{figure}
\begin{center}
\includegraphics[width=8cm \vspace{-0.3cm}]{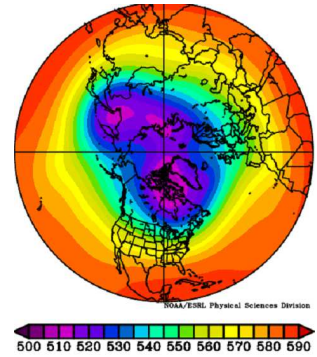}\vspace{-5cm}
\caption{\label{fig2} Northern Hemisphere, Jan-Apr 2018y:  500-hPa/mb Height 90 day atmospheric pressure mean (at approximately 5500 meters)  \cite{psd}.} 
\end{center}
\end{figure}
\begin{figure}
\begin{center}
\includegraphics[width=8cm \vspace{-0.3cm}]{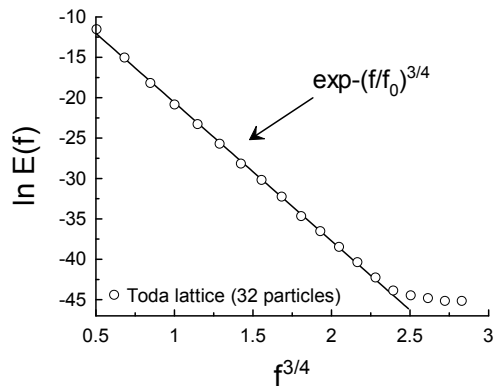}\vspace{-4.2cm}
\caption{\label{fig3} The power spectrum (a high-frequency part) of the Toda particle coordinate $q_1$ fluctuations for the direct numerical simulation reported in the Ref. \cite{ebf}. } 
\end{center}
\end{figure}
  
\begin{figure}
\begin{center}
\includegraphics[width=8cm \vspace{-0.25cm}]{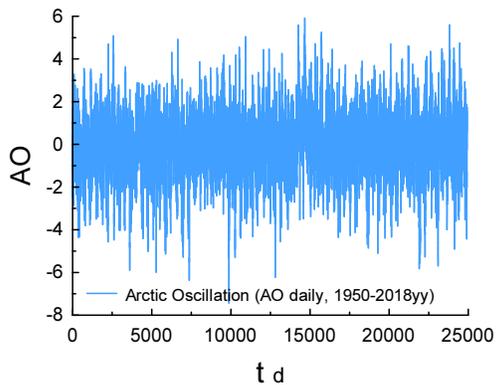}\vspace{-3.97cm}
\caption{\label{fig4} The AO daily index since 1950y. } 
\end{center}
\end{figure}

  The AO is naturally related to the North Atlantic Oscilation (NAO) \cite{hur}-\cite{far}, which is known to be chaotic \cite{b1}. However, the AO index dynamics is rather different from that of the NAO index. The question is: Whether the annular modes are chaotic (deterministic) or stochastic \cite{hir}? \\

 Exponential power spectra are typical for chaotic systems with broadband spectrum (see, for instance \cite{fm}-\cite{sig}):
$$
 E(f) \propto \exp-(f/f_c) \eqno{(1)}
$$ 
  However, for Hamiltonian dynamical systems a more complex situation takes place. Corresponding power spectra are stretched exponentials, which come from a weighted superposition of the exponentials 
$$
E(f ) \propto \int_0^{\infty} P(f_c) \exp -(f/f_c)~ df_c  \propto \exp-(f/f_0)^{3/4}  \eqno{(2)}
$$
This is the Hamiltonian distributed chaos \cite{b2}.  

  Figure 3 shows, as an example, a high-frequency part of the power spectrum for the coordinate $q_1$ of a particle in the classic Toda lattice \cite{toda}. The Toda lattice is a one dimensional Hamiltonian dynamical system with $N$ particles. The particles having the coordinates $q_i$ interact with each other with the potential $V(q_i,q_{i+1})=\ e^{-(q_i-q_{i+1})}$. The dynamic equations for this system are:
$$
\dot{q_i}=p_i,   \eqno{(3)}
$$
$$
\dot{p_i}=e^{q_i-q_{i-1}}-e^{q_{i+1}-q_i}  \eqno{(4)} 
$$
 The spectral data for the Fig. 3 were taken from a direct numerical simulation (with random initial conditions, periodic boundary conditions - on a ring, and zero total momentum) reported in the Ref. \cite{ebf}. The straight line in the Fig. 3 is drawn in order to indicate correspondence to the Eq. (2) in the appropriately chosen scales. \\

 \begin{figure}
\begin{center}
\includegraphics[width=8cm \vspace{-0.42cm}]{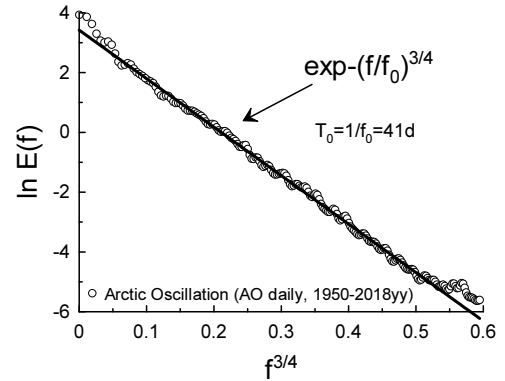}\vspace{-3.8cm}
\caption{\label{fig5} The power spectrum of the AO daily index. } 
\end{center}
\end{figure}

In geophysical fluid dynamics most of the relevant theoretical models are Hamiltonian \cite{she}-\cite{gl2}. Therefore, one can expect that the Hamiltonian distributed chaos can be relevant also to the Arctic and Antarctic Oscillations. Figure 4 shows the daily AO index since January 1950y \cite{ao}. Figure 5 shows corresponding power spectrum calculated with the maximum entropy method, that provides an optimal resolution for the short data sets \cite{oh}. The straight line indicates correspondence to the Eq. (2). The characteristic time scale inferred from the best fit is $T_0=1/f_0 \simeq 41d$. The near 40-day oscillations are well known for the  extratropics and are usually associated with the interaction of the non-zonal flows with topography (see, for instance, \cite{b1},\cite{mag}-\cite{cun} and references therein).  Figure 6 shows power spectrum corresponding to the AAO daily index \cite{ao} since January 1979y. The straight line indicates correspondence to the Eq. (2). The characteristic time scale inferred from the best fit is $T_0=1/f_0 \simeq 41d$ as for the AO daily index (cf. Fig. 5).   \\

  A remarkable apparent similarity of the annular modes of the atmospheric circulation - AO and AAO, was already mentioned for the mean meridional circulations, for zonal wind fields and for many other structural features \cite{tw}. Comparison of the Fig. 5 with the Fig. 6 provides a new quantitative information about this similarity.

\begin{figure}
\begin{center}
\includegraphics[width=8cm \vspace{-1cm}]{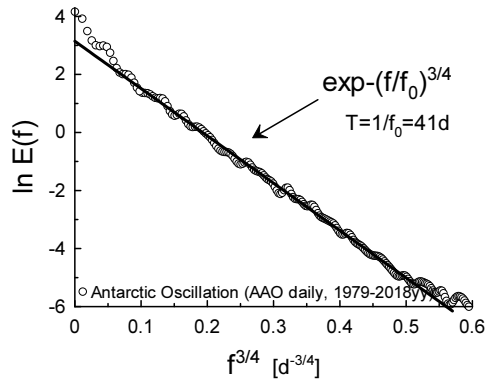}\vspace{-3.7cm}
\caption{\label{fig6} The power spectrum of the AAO daily index. } 
\end{center}
\end{figure}

\section{Acknowledgement}

I acknowledge use of the data provided by the NOAA: the Earth system research laboratory (Physical Sciences Division) and the Climate Prediction Center.

\end{document}